\documentstyle[preprint,aps,epsfig]{revtex}
        \def\be{\begin{equation}}
        \def\ee{\end{equation}}

\begin{document}
\begin{titlepage}
\vspace*{5mm}

\begin{center} {\large \bf A Two-Species Exclusion Model 
With Open Boundaries: A use of $q$-deformed algebra} \\

\vskip 1cm

\centerline {\bf Farhad H Jafarpour \footnote {e-mail:JAFAR@theory.ipm.ac.ir}} 
\vskip 1cm
{\it  Department of Physics, Sharif University of Technology, }\\
{\it P.O.Box 11365-9161, Tehran, Iran }\\
{\it  Institute for Studies in Theoretical Physics and Mathematics,}\\
{\it P.O.Box 19395-5531, Tehran, Iran}
\end{center}

\begin{abstract}
In this paper we study an one-dimensional two-species exclusion model
with open boundaries. The model consists of two types of particles moving
in opposite directions on an open lattice. Two adjacent particles swap
their positions with rate $p$ and at the same time they can return to
their initial positions with rate $q$ if they belong to the different
types. Using the Matrix Product Ansatz (MPA)
formalism, we obtain the exact phase diagram of this model in restricted
regions of its parameter space. It turns out that the model has two
distinct phases in each region. We also obtain the exact expression for
the current of particles in each phase. \\

\end{abstract}



\end{titlepage}

\newpage

\section{Introduction}

The stationary state properties of one-dimensional driven diffusive
systems are currently of much research interest [1-5]. These systems
exhibit very interesting cooperative phenomena such as boundary-induced
phase transitions,
spontaneous symmetry breaking and single-defect induced phase transitions
which are absent in one-dimensional equilibrium statistical mechanics. Many 
physical phenomena such as hopping conductivity, growth processes and traffic 
flows can also be explained by these models [6-9]. One of the most basic
model is the Asymmetric Simple Exclusion Process (ASEP), which shows a rich
behavior [10]. 
The ASEP is a model of particles diffusing on a lattice driven by an external 
field and
with hard-core exclusion. Models with more than one kind of particles have 
also been investigated. The ASEP in the presence of a second class particle 
(impurity) has provided a framework for the study of shocks (see [5] and 
[9]). Another model of this kind is the Partially Asymmetric Simple
Exclusion Process (PASEP). In this model, particles are allowed to hop
both to their immediate right and left sites with unequal rates. This
model has been studied both with open boundaries [20] and in the presence
of an impurity on a ring [11].   
A multi-species ASEP has also been suggested, which seems to be a simple
realization for real traffic [12]. \\
In this paper we consider a model containing two types of particles on a 
lattice of length $L$ with open boundary condition. The two types of particles,
which we refer to them as " positive " and " negative " particles, move in
opposite direction. The positive (negative) particles are injected (removed)
from the left-most site of the lattice and are removed (injected) from the 
right-most site of the lattice. Every where through the lattice, two
adjacent particels
interchange their positions, unless they are both positive or negative.
The system evolves according to an stochastic 
dynamical rule as follows. In each infinitesimal time step $dt$ the following
events occur at each nearest-neighbour pair of sites 
$i, i+1$ $(1 \le i \le L-1)$:\\
\begin{eqnarray}
(+)(0) \rightarrow (0)(+) \ \ \  with \ \ \  rate \ \ \  1  \nonumber \\
(0)(-) \rightarrow (-)(0) \ \ \  with \ \ \  rate \ \ \  1   \\
(+)(-) \rightarrow (-)(+) \ \ \  with \ \ \  rate \ \ \  p \nonumber \\
(-)(+) \rightarrow (+)(-) \ \ \  with \ \ \  rate \ \ \  q  \nonumber  
\end{eqnarray}
where $(+)$ and $(-)$ indicate a positive or a negative particle,
respectively, and $(0)$ indicates an empty site.
Also, in each infinitesimal time step $dt$, the following events may occur 
at the first $(i=1)$ and the last $(i=L)$ site of the lattice: \\
\begin{eqnarray}
At \ \ site \ \ & i=1 & \ \ \left\{
  \begin{array}{ll}
  (0) \rightarrow (+) \ \ \  with \ \ \  rate \ \ \  \alpha  \nonumber \\
  (-) \rightarrow (0) \ \ \  with \ \ \  rate \ \ \  \beta  \nonumber
  \end{array}\right.
  \\
At \ \ site \ \ & i=L & \ \ \left\{
  \begin{array}{ll}
  (0) \rightarrow (-) \ \ \  with \ \ \  rate \ \ \  \alpha   \\
  (+) \rightarrow (0) \ \ \  with \ \ \  rate \ \ \  \beta 
   \end{array}\right.
   \end{eqnarray}
For $q=0$, this model reduces to the model introduced in [13,14] 
which using simulation data and doing exact calculations has been
extinsively studied. These authors  
have shown that for certain values of the parameters $\alpha, \beta$ and
$p$
the symmetry of dynamics under interchange of positive and negative particles
and of their directions is spontaneously broken. \\
In reference [15] Alcaraz et al have studied the $N$-species stochastic
models with open boundary and found their related algebras, which appear
in
the MPA formalism first introduced in [16]. Our model can be considered as
a $N=2$ case which will be studied in details.\\
The process $(1)$ has also been considered on a closed ring in [17-19]. It
has been shown that when the density of positive particles is equal to
the density of negative ones, depending on the values of the parameters
of the model, three phases exist: a pure
phase in which one has three pinned blocks of only positive, negative particels 
and vacancies (where the translational invariance is spontaneously
broken); 
a mixed phase with a non-vanishing current of particles; and a disordered
phase.
Here we study the effects of the open boundaries. For certain cases $(\beta=1$
or $\alpha=\infty)$ we are able to solve our model exactly and find the
modified phase diagrams (in comparison with $q=0$ case). We will show that
for $\alpha=\infty$, where the system is devoid of vacancies, only two phases
exist. In $\beta=1$ limit the model has also two distinct phases in which the 
current of positive particles is equal to those of negative ones.\\ 
This paper is organized as follows. In section $2$ we will present the
exact solution of the model for the case $ \alpha=\infty$ using the known
results. We will also obtain the exact generating function of the 
partition
function of the model using the Matrix Product Ansatz (MPA) and calculate
the current of the particles in $\beta=1$ limit. In the last section we will 
compare our results with those obtained in [13] for $q=0$.\\ 
\section{Matrix Product Solutions}

In this section we will show that the stationary probability of the model 
defined in $(1)$ and $(2)$ can be obtained using the MPA 
for two specific cases $\alpha=\infty$ and $\beta=1$. According to the MPA
formalism, the stationary probability $P(\{ C \})$ of any configuratin $\{C \}$
can be written as a matrix element of a product of non-commuting
operators.
Before reviewing this approach we define some notations. We introduce two
occupation numbers, $\tau_i$ and $\theta_i$, for each site $i$, where $\tau_i=1$ 
if site $i$ is occupied by a positive particle and $\tau_i=0$ otherwise.
Similarly, $\theta_i=1$ if site $i$ is occupied by a negative particle and 
$\theta_i=0$ otherwise. Since the process is exclusive, so that each site of the
lattice can only be occupied at most by one particle, each configuration of
the system is uniquely defined by the set of occupation numbers 
$\{ \tau_i,\theta_i \}$. Now the normalized stationay state weight for a 
lattice of size $L$ can be witten as:\\
\be
P(\{ \tau_i,\theta_i \})=
{1 \over Z_L} \langle W \vert \prod_{i=1}^{L} \{ \tau_i D + \theta_i A
+(1-\tau_i-\theta_i) E \} \vert V \rangle.
\ee
The normalization factor $Z_L$ in the denominator of the equation $(3)$,
which plays a role analogous to the partition function in equilibrium 
statistical mechanics, is a fundumental quantity and can be calculated using 
the fact $ \sum_{ \{ \tau_i,\theta_i \} } P(\{ \tau_i,\theta_i \})=1$. Thus
one finds\\
\be
Z_L = \sum_{ \{ \tau_i,\theta_i \} }  
\langle W \vert \prod_{i=1}^{L} \{ \tau_i D + \theta_i A
+(1-\tau_i-\theta_i) E \} \vert V \rangle=\langle W \vert G^{L}\vert V \rangle
\ee 
in which $G=D+A+E$. The operators $D$, $A$ and $E$ correspond to the presence
of a positive, a negative particle, and a hole respectively. These
operators with the vectors 
$\vert V \rangle$ and $\langle W \vert$ satisfy a certain algebra which will be 
discussed below.\\
\subsection{The limit $\alpha \rightarrow \infty$}
In this limit, as soon as a hole appears at a boundary site, it is
removed. Therefore in the steady state the lattice will be empty of
holes. Now the dynamical rules given by $(1)$ and $(2)$ reduce to\\ 
\begin{eqnarray}
(+)(-)  \rightarrow  (-)(+) \ \ \  with \ \ \  rate \ \ \ p \nonumber \\
(-)(+)  \rightarrow  (+)(-) \ \ \  with \ \ \  rate \ \ \ q   \\
At \ \ site \ \ i=1 \ \ (-)  \rightarrow  (+) \ \ \  with \ \ \  rate \ \
\ \beta \nonumber \\
At \ \ site \ \ i=L \ \ (+)  \rightarrow  (-) \ \ \  with \ \ \  rate \ \
\ \beta \nonumber        
\end{eqnarray}
Using the MPA one obtains the following quadratic algebra for this case\\ 
\begin{eqnarray}
pDA-qAD & = & D+A \nonumber \\
{\beta} \langle W \vert A & = & \langle W \vert \\
{\beta} D \vert V \rangle & = &  \vert V \rangle. \nonumber 
\end{eqnarray}
Now if one imagine the negative particles as holes, the problem reduces 
to the single-species PASEP
with open boundaries and equal injection and extraction rates. 
As we mentioned, recently 
the PASEP has been studies widely with open boundaries [20]. Using
the results given there, we find two following phases in
the thermodynamic limit $(L \rightarrow \infty)$:\\

$I. \ \ { \frac{\beta}{p-q}>{1 \over 2}} $ \\

The current of the positive particels $J_{+}$ is equal to the current of the 
negative ones $J_{-}$ and has its maximum value\\
\be
J_{+}=J_{-}=\frac{p-q}{4}.
\ee
Also the density of the positive particles $<\tau_i>$ in the bulk has a 
power law behavior\\
\be
<\tau_i> \simeq {\frac{1}{2}} + {\frac{1}{2 \pi i^{1/2} }}.
\ee
The density of the negative particels $<\theta_i>$ can be obtained using the 
equality $<\tau_i>+<\theta_i>=1$.\\

$II. \ \ { \frac{\beta}{p-q}<{1 \over 2}} $  \\

The current of the positive particels is again equal to the
current of the negative ones and is given by\\
\be
J_{+}=J_{-}=\beta(1- \frac{\beta}{p-q}).
\ee
The density profile of the positive particles is linear in the bulk which is
a consequence of the superposition of shocks\\
\be
<\tau_i> \simeq {\frac{\beta}{p-q}} + i({1-2 \frac{\beta}{p-q}}).
\ee
This phenomenon has also been observed in the ASEP with open boundaries
when the injection and extraction rates become equal [10].
\subsection{The limit $\beta=1$}

Another limit which can be solved using the MPA formalism exactly is
$\beta=1$.
The operators and vectors satisfy the following algebra\\
\begin{eqnarray}
pDA-qAD & = & \alpha(D+A) \nonumber \\
DE & = & \alpha E \nonumber \\
EA & = & \alpha E \nonumber \\
E \vert V \rangle & = & \vert V \rangle \\
D \vert V \rangle & = & \alpha \vert V \rangle \nonumber \\
\langle W \vert E & = & \langle W \vert \nonumber \\
\langle W \vert A & = & \alpha \langle W \vert \nonumber .
\end{eqnarray}
Following [13] one can choose\\
\be
E=\vert V \rangle \langle W \vert \ \ \ , \ \ \ \langle V \vert W \rangle =1. 
\ee
Then the algebra $(11)$ can be written as\\
\begin{eqnarray}
(\frac{p}{\alpha})DA-(\frac{q}{\alpha}) AD & = & D+A \nonumber \\
D \vert V \rangle & = & \alpha \vert V \rangle \\
\langle W \vert A & = & \alpha \langle W \vert \nonumber .
\end{eqnarray}
This algebra is very similar to the algebra associated with the PASEP [19,20]. Here
we adopt the same representation proposed in [19]. One can easily check
that the following representation satisfy $(12)$ and $(13)$\\
\be 
{D} ={\alpha \over p-q} 
\left( \begin{array}{llllll} 
1+a&\sqrt{c_1}&0&0&.&.\\
0&1+a(\frac{q}{p})&\sqrt{c_2}&0&.&.\\
0&0&1+a(\frac{q}{p})^2 & \sqrt{c_3}&.&.\\
0&0&0&1+a(\frac{q}{p})^3 &. &.\\
.&.&.&.&.&.\\
.&.&.&.&.&.\\ 
\end{array} \right ) \ , \ A=D^{T} \ , 
\ E=\vert 0 \rangle \langle 0 \vert. 
\ee 
Where the superscript $T$ indicates the transpose,\\
\be
c_n=(1-({\frac{q}{p}})^n)(1-a^2 ({\frac{q}{p}})^{n-1}) \ \ , \ \ a=p-q-1, 
\ee
$\langle 0 \vert =[1 0 0 0....] $ and $\vert 0 \rangle = \langle 0 \vert^{T}$.
Using the matrix algebra given by $(12)$ and $(13)$ we find the following  
expressions for the current of the positive and the negative particles in the
stationary state\\
\begin{eqnarray}
J_{+}= 
\frac {\langle W \vert G^{i-1} (pDA-qAD+DE) G^{L-i-1} \vert V \rangle}
{\langle W \vert G^L \vert V \rangle} 
      = \alpha 
\frac{\langle W \vert G^{L-1} \vert V \rangle} 
{\langle W \vert G^{L} \vert V \rangle}, \nonumber \\
J_{-}=
\frac{\langle W \vert G^{i-1}(pDA-qAD+EA) G^{L-i-1} \vert V \rangle}
{\langle W \vert G^L \vert V \rangle} 
     = \alpha 
\frac{\langle W \vert G^{L-1} \vert V \rangle} 
{\langle W \vert G^{L} \vert V \rangle}.
\end{eqnarray}
As can be seen from $(16)$, the currents are site-independent (as it
should be
in the stationary state), equal and given by the matrix element of powers of $G$.
In what follows we will introduce a generating function to calculate the matrix
element of all powers of $G$ betwen the vectors $\vert V \rangle $ and 
$\langle W \vert $.\\
Define a generating function\\
\be
f(\lambda):=\sum_{L=1}^{\infty} 
\lambda^{L-1} \langle W \vert G^L \vert V \rangle.
\ee
The convergence radius of this formal series, $R$, is proportional to the 
current of particles given by $(16)$ in the thermodynamic limit\\
\be
R= lim_{L \rightarrow \infty}  
{ \frac{\langle W \vert G^{L-1} \vert V \rangle} 
{\langle W \vert G^{L} \vert V \rangle} }.
\ee
On the other hand, the radius of convergence is the absolute value of the
nearest 
singularity of $f(\lambda)$ to the origin. Once we obtain the singularities
of the function $f(\lambda)$, we can calculate the current of particles and 
distinguish the phases.\\
Using $(12)$ and $(13)$ one can expand the expression 
$\langle W \vert G^L \vert V \rangle$ as\\ 
\be
\langle G^L \rangle := \langle W \vert G^L \vert V \rangle=
\sum_{r\ge0}\hskip-30pt\sum_{
\begin{array}{c}
\scriptstyle j_0,j_r \geq 0 \\\noalign{\vskip-5pt}
\scriptstyle j_1,m_1,\ldots,j_{r-1},m_r >0\\\noalign{\vskip-4pt}
\scriptstyle j_0+m_1+j_1+\cdots + m_r+j_r = L
\end{array}}
\langle W \vert C^{j_0} E^{m_1} C^{j_1} \cdots E^{m_r} C^{j_r} \vert V \rangle
\ee
where $C:=D+A$. Noting that $E^m=E$, after some computation, we obtain\\
\be
\langle G^L \rangle = \langle C^L \rangle+
\sum_{r=1}^L\hskip-30pt\sum_{
\begin{array}{c}
\scriptstyle j_0,j_r \geq 0 \\\noalign{\vskip-5pt}
\scriptstyle j_1,\ldots,j_{r-1}>0\\\noalign{\vskip-4pt}
\scriptstyle j_0+j_1+\cdots +j_r \le L-r
\end{array}}
\frac{(L-1-j_0-\ldots-j_r)!}{(r-1)!(L-r-j_0-\ldots-j_r)!} 
\langle C^{j_0} \rangle \ldots \langle C^{j_r} \rangle
\ee
and also\\
\be
f(\lambda)=\sum_{L=1}^{\infty} \lambda^{L-1} \langle G^L \rangle=
\frac { (1-\lambda^{-1})+\lambda^{-2}(1+\lambda)
\sum_{n=0}^{\infty} \lambda^{n+1} \langle C^n \rangle}
{1-\sum_{n=0}^{\infty} \lambda^{n+1} \langle C^n \rangle}.
\ee
It is known that the expression 
$g(\lambda):=\sum_{n=0}^{\infty}\lambda^{n+1} \langle C^n \rangle$ can be written
explicitly in terms of the basic $q$-hypergeometric function (see [19]
and references therein)\\
\be
g(x(\lambda))= x \frac{(p-q)}{\alpha}
  \frac{( \frac{q}{p} x^2; 
\frac{q}{p})_{\infty}( \frac{q}{p};
\frac{q}{p})_{\infty}}{(ax; \frac{q}{p})_{\infty}^2} {_2\phi_1}
\left[ \begin{array}{cc}  
               ax,ax  \\
               \frac{q}{p}x^2       
         \end{array};
               \frac{q}{p}, \frac{q}{p} 
  \right]
\ee
in which $x(\lambda)={\frac{1}{2}} 
\{ \frac{p-q}{\lambda \alpha} -2-
\sqrt{ 
({\frac{p-q}{\lambda \alpha}})^2 
-4
{\frac{p-q}{\lambda \alpha}}  } 
\}$. The quantities $(z;q)_n$ and $(z;q)_{\infty}$ are defined as\\
\begin{eqnarray}
(z;q)_n & = & \left\{
  \begin{array}{ll}1, \ \ if \ \ \ n=0,\\
  (1-z)(1-zq)(1-zq^2)\cdots(1-zq^{n-1}), \ \ if \ \ \ n>0 ,
   \end{array}\right.
  \\
  (z;q)_{\infty} & =& \prod_{n=0}^{\infty} (1-zq^n) \nonumber.
\end{eqnarray}
Lastly, the basic $q$-hypergeometric function is defined by the series\\
\begin{equation}
  \label{b_hyp}
  {_2\phi_1}
  \left[ \begin{array}{cc}  
               a_1,a_2  \\
               b       
         \end{array};
                        q ,z 
  \right]
  =
  \sum_{n=0}^{\infty}
  \frac{(a_1;q)_n (a_2;q)_n  }
       {(b;q)_n (q;q)_{n}} 
  z^n
\end{equation}
which tends to the usual hypergeometric series as $q \rightarrow 1$. The
$_2\phi_1$ series converges when $0<|q|<1$ and $|z|<1$ [20]. In this paper
the convergence condition of $(22)$ is $q<p$; therefore, without losing
the generality, we limit ourselves to this region.\\
As we mentioned, the singularities of $f(\lambda)$ specify the phase diagram 
of the model. From the expression $(21)$, we see that there are two possible 
sources for the singularities: the singularities of $g(x(\lambda))$ and a zero 
of the denominator of $(21)$. First we consider the singularities of 
$g(x(\lambda))$. From $(22)$ one can see that $g(x(\lambda))$ has two 
singularities:
$\lambda_1=\frac{p-q}{4\alpha}$ which is a square root singularity and 
$\lambda_2=\frac{p-q-1}{\alpha(p-q)}$ which is a simple root. In order to
discuss the zeros of the denominator of $f(\lambda)$, we use the same assumption
proposed in [19]. In the convergence region of $(22)$ i.e. $q<p$, for 
$0 \leq x \leq 1$ the function $g(x)$ satisfies $g^{'}(x)>0$. It means that the
function $g(x)$ increases monotonically from $0$ to $g(1)$ when $0 \leq x
\leq 1$;
therefore, the equation $g(x(\lambda_3))=1$ (which gives the zeros of the
denominator
of $(21)$) has only one root in this region. Comparing the absolute value of the 
singularities $\lambda_1$, $\lambda_2$ and $\lambda_3$, one can easily
find the following results:\\

$I)$ For $p-q \geq 2$ the radius of convergence $(18)$ of the formal
series $(17)$ is equal to $R=\lambda_3<\lambda_1,\lambda_2$ which is the 
solution of the equation $g(x(\lambda_3))=1$. 
Since $\lambda_3$ is a simple pole, we expect that $Z_{L}$ behaves
asymptotically $(L \rightarrow \infty)$ as $\lambda_3^{-L}$. 
The current of the particles $(16)$ can also be obtained\\
\be
J_{+}=J_{-}=\alpha\lambda_3
\ee

$II)$ For $p-q<2$ two different situations may occur. 
In the region specified by $g(1)\leq1$ and $p-q<2$, we find
$R=\lambda_1=\frac{p-q}{4\alpha}$ and the current of particles to be\\
\be
J_{+}=J_{-}=\frac{p-q}{4}
\ee
In the region $g(1)>1$, it turns out that $R=\lambda_3<\lambda_1,\lambda_2$
which is again the solution of the equation $g(x(\lambda_3))=1$. The
partition function $Z_{L}$ again behaves as $\lambda_3^{-L}$ and the current 
of particles in this case can be obtained from $(25)$. The boundary of
these two recent phases will be specified by\\
\be
g(1)= \frac{(p-q)}{\alpha}
  \frac { (\frac{q}{p};\frac{q}{p})_{\infty}^2}
{(p-q-1;\frac{q}{p})_{\infty}^2}
{_2\phi_1}
\left[ \begin{array}{cc}  
               p-q-1,p-q-1  \\
               \frac{q}{p}       
         \end{array};
                        \frac{q}{p},\frac{q}{p} 
  \right]=1
\ee
In Fig.1 we have plotted the phase diagram of our model in $\beta=1$
limit both for $q \neq 0$ (the left diagram) and $q=0$ (the right
diagram). As we mentioned in section (2) for $\alpha=\infty$, the line
$p-q \geq 2 $ is the line of shock configurations. The bold lines in Fig.1
mark these lines.

\begin{figure}
\centerline{\epsfig{figure=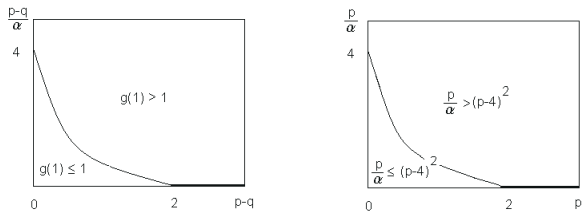,width=.5\columnwidth}}
\vspace{1 mm}
\caption{ Plot of the phase diagrams for $q=0$ (the right diagram) and
$q\neq 0$ (the left diagram) in $\beta=1$ limit. }
\end{figure}
\section{Comparison and Concluding Remarks}

In this paper we studied a generalized two-species exclusion model
with open boundaries. The positive particles are supplied at the left end 
of the chain and they leave it at the right end. Similarly, the negative 
particles are supplied at the right end and they leave the system at the left
end. As soon as a positive and a negative particle meet each other (the positive
particle is supposed to be in the left hand side of the negative
particle), they
interchange their positions with rate $p$. At the same time they may go to their
initial positions with rate $q$. In $q=0$ limit this model reduces to the
one studied in [13]. Using the MAP formalism, our model has been
studied in two
different limits ($\alpha=\infty$ and $\beta=1$) and the corresponding phase
diagrams obtained. It has been shown that all the phases are symmetric in which 
the current of the positive and negative particles are equal. One can easily
check that for $q=0$ all the results obtained here reduce to those obtained 
in [13]. In comparison with [13], as can be seen in Fig.1, the phase
diagram of the model has been modified.\\
The study of the whole parameters space of the model proposed in this paper
is still an open problem. Using the mean field approximation and simulation 
data,
the authors have shown that in $q=0$ limit this model has also asymmetric 
phases where the current of the positive and negative particles become  
different. It will be interesting to study the structure of asymmetric phases
of this model which may occur for the certain values of the parameters $\alpha$,
$\beta$, $p$ and $q$.\\

{ \large \bf Acknowledgement: } \\
I would like to thank V. Karimipour for reading the manuscript.


\begin{thebibliography}{99}
\def\ll #1 #2 #3{{\bf #1} (19#2) #3}
\def\nm #1 #2 {\bibitem {#1} #2}
\def\gr #1 {{\large #1}}
\def\PRA{{\em Phys.\ Rev.\ }{\bf A\ }} 
\def\PRB{{\em Phys.\ Rev.\ }{\bf B\ }}
\def\PRE{{\em Phys.\ Rev.\ }{\bf E\ }}
\def\PRL{{\em Phys.\ Rev.\ Lett.\ }}
\def\JPA{{\em J.\ Phys.\ A: Math\ Gen.\ }}
\def\JPC{{\em J.\ Phys.\ C: Solid State Phys.\ }}
\def\JP{{\em J.~Physique }}
\def\JPI{{\em J.\ Phys.\ I France\ }}
\def\EPL{{\em Europhys.\ Lett.\ }}
\def\ZPB{{\em Z.\ Phys.\ }{\bf B}}
\def\JSP{{\em J.\ Stat.\ Phys.\ }}
\def\JJpn{{\em J.\ Phys.\ Soc.\ Jpn.\ }}



\bibitem{ligget1} T. M.\ Ligget,
 Intracting Particle Systems (Springer-Verlag, New York, 1985).


\bibitem{ligget2} T. M.\ Ligget,
 Stochastic Intracting Systems: Contact, Voter, and Exclusion Processes
 (Springer-Verlag, New York, 1999).


\bibitem{spohn} H.\ Spohn,
 Large Scale Dynamics of Intracting Particles (Springer-Verlag, New
 York,1991).
        


\bibitem{zia} B.\ Schmittmann and R. K. P.\ Zia,
 Statistical mechanics of driven diffusive systems, in Phase Transitions
 and Critical Phenomena, Vol 17, C.\ Domb and J.\ Lebowitz eds. (Academic,
 London, 1994).     


\bibitem{mallick} K.\ Mallick, 
 Shocks in the asymmetry exclusion model with an impurity,
 \JPA{\bf 29}, 5375 (1996).


\bibitem{Richards} P.M.\ Richards, Theory of one-dimensional hopping 
 conductivity and diffusion, \PRB{\bf 16}, 1393 (1977).



\bibitem{Nagel} K.\ Nagel and M.\ Schreckenberg, \JPI {\bf 2}, 2221
 (1992). 



\bibitem{Krug} J.\ Krug, Boundary induced phase transition in driven 
 diffusive systems, \PRL{\bf 67}, 1882 (1991). 



\bibitem{lee} H-W.\ Lee, V.\ Popkov and D.\ Kim, Two way traffic flow:
 Exactly solvable model of traffic jam, \JPA{\bf 30}, 8497 (1997).



\bibitem{Derrida} B.\ Derrida, An exactly soluble non-equilibrium system: 
 The asymmetric simple exclusion model, 
 {\em Phys.\ Rep.\ } { \bf 301}, 65 (1998). 


\bibitem{sasajafar} F. H.\ Jafarpour, Partially asymmetric simple
 exclusion model in the presence of an impurity on a ring,  \JPA{\bf 33},
 1797 (2000), \\   
 T.\ Sasamoto, 
 One-dimensional partially asymmetric simple exclusion process on a ring
 with a defect particle, cond-mat/9910483  and also [5].
 


\bibitem{karimipour} V.\ Karimipour,
 A multi-species asymmetric simple exclusion process and its relation to
 traffic flow \PRE{\bf 59}, 205 (1999).
 


\bibitem{Evansetal} M. R.\ Evans, D.P.\ Foster, C.\ Godreche and
 D.\ Mukamel, Asymmetric exclusion model with two species: Spontaneous
 symmetry breaking, \JSP{\bf 80}, 69 (1995) and  \\ 
 M. R.\ Evans, D.P.\ Foster, C.\ Godreche and D.\ Mukamel, Spontaneous
 symmetry breaking in  one dimensional driven diffusive system, \PRL{\bf74}, 
 208 (1995).
  

\bibitem{rittenbetal}  P.F.\ Arndet, T.\ Heinzel and V.\ Rittenberg,  
 First-order phase transitions in one-dimensional steady states,
 \JSP{\bf 90}, 783 (1998).




\bibitem{alcarazetal} F. C.\ Alcaraz, S.\ Dasmahapatra and V.\ Rittenberg,
 N-species stichastic models with boundaries and quadratic algebras,
 \JPA{\bf 31}, 845 (1998).



\bibitem{derrida93} B.\ Derrida, M.R.\ Evans, V.\ Hakim and V.\ Pasquier, 
 Exact solution of a 1d asymmetric exclusion model using a matrix formulation,
 \JPA{\bf 26}, 1493 (1993).




\bibitem{Rittenbergetal} P.F.\ Arndet, T.\ Heinzel and V.\ Rittenberg,
 Spontanious breaking of translational invariance and spatial condensation
 in stationary states on a ring, \JSP{\bf 97}, 1 (1999).



\bibitem{Rittenbergetal} P.F.\ Arndet, T.\ Heinzel and V.\ Rittenberg,
 Spontaneous breaking of translational invariance in one-dimensional
 stationary states on a ring: The neutral system, \JPA{\bf 31}, L45
 (1998).


\bibitem{rajewskyetal} N.\ Rajewsky, T.\ Sasamoto and E. R.\ Speer,
 Spatial particle condensation for an exclusion process on a ring,
 cond-mat/9911322.
      


\bibitem{sasamoto} T.\ Sasamoto, One-dimensional partially asymmetric
 simple exclusion model with open boundaries: Orthogonal polynomials
 approach, \JPA{\bf 32}, 7109 (1999).
  


\end{thebibliography}
\end{document}